\documentclass[sigconf]{acmart}
\renewcommand\footnotetextcopyrightpermission[1]{} 

\AtBeginDocument{%
  }

\usepackage{orcidlink}
\usepackage{graphicx}
\newcommand{\rotcell}[1]{%
  \makebox[2pt][r]{\rotatebox[origin=r]{-35}{#1}}%
}

\newcommand{\unitdegree}[1]{{#1}°}

\newcommand{\unitcmeter}[1]{{#1}\,cm}

\newcommand{\unitsecond}[1]{{#1}\,sec}
\newcommand{\unitminute}[1]{{#1}\,min}

\newcommand{\unithertz}[1]{{#1}\,Hz}

\newcommand{\unitpercent}[1]{{#1}\,\%}

\newcommand{\enquote}[1]{``{#1}''}
\begin{document}


\title{Unobtrusive In-Situ Measurement of Behavior Change by Deep Metric Similarity Learning of Motion Patterns}

\renewcommand{\shorttitle}{Unobtrusive In-Situ Measurement of Behavior Change by Similarity Learning of Motion Patterns}

\author{Christian Merz\orcidlink{0000-0001-6655-416X} }
\authornote{Both authors contributed equally to this research.}
\email{christian.merz@uni-wuerzburg.de}
\orcid{0000-0001-6655-416X}
\affiliation{%
  \institution{University of Wurzburg}
  \country{Germany}}

\author{Lukas Schach\orcidlink{0009-0002-1876-1994} }
\orcid{0009-0002-1876-1994}
\authornotemark[1]
\email{lukas.schach@uni-wuerzburg.de}
\affiliation{%
  \institution{University of Wurzburg}
  \country{Germany}}

\author{Marie Luisa Fiedler\orcidlink{0000-0002-3472-3798}}
\orcid{0000-0002-3472-3798}
\email{marie.fiedler@uni-wuerzburg.de}
\affiliation{%
  \institution{University of Wurzburg}
  \country{Germany}}
  
\author{Jean-Luc Lugrin\orcidlink{0000-0002-2725-2123}}
\orcid{0000-0002-2725-2123}
\email{jean-luc.lugrin@uni-wuerzburg.de}
\affiliation{%
  \institution{University of Wurzburg}
  \country{Germany}}
  
\author{Carolin Wienrich\orcidlink{0000-0003-3052-7172}}
\orcid{0000-0003-3052-7172}
\email{carolin.wienrich@uni-wuerzburg.de}
\affiliation{%
  \institution{University of Wurzburg}
  \country{Germany}}

\author{Marc Erich Latoschik\orcidlink{0000-0002-9340-9600}}
\orcid{0000-0002-9340-9600}
\email{marc.latoschik@uni-wuerzburg.de}
\affiliation{%
  \institution{University of Wurzburg}
  \country{Germany}}
\renewcommand{\shortauthors}{Trovato et al.}

\begin{abstract}
  This paper introduces an unobtrusive in-situ measurement method to detect user behavior changes during arbitrary exposures in XR systems. Here, such behavior changes are typically associated with the Proteus effect or bodily affordances elicited by different avatars that the users embody in XR. We present a biometric user model based on deep metric similarity learning, which uses high-dimensional embeddings as reference vectors to identify behavior changes of individual users. We evaluate our model against two alternative approaches: a (non-learned) motion analysis based on central tendencies of movement patterns and subjective post-exposure embodiment questionnaires frequently used in various XR exposures. In a within-subject study, participants performed a fruit collection task while embodying avatars of different body heights (short, actual-height, and tall). Subjective assessments confirmed the effective manipulation of perceived body schema, while the (non-learned) objective analyses of head and hand movements revealed significant differences across conditions. Our similarity learning model trained on the motion data successfully identified the elicited behavior change for various query and reference data pairings of the avatar conditions. The approach has several advantages in comparison to existing methods: 1) In-situ measurement without additional user input, 2) generalizable and scalable motion analysis for various use cases, 3) user-specific analysis on the individual level, 4) with a trained model, users can be added and evaluated in real time to study how avatar changes affect behavior.
\end{abstract}

\ccsdesc[500]{Computing methodologies~Cross-validation}
\ccsdesc[300]{Human-centered computing~Virtual reality}
\ccsdesc[100]{Human-centered computing~Empirical studies in HCI}

\keywords{Virtual reality, Motion data, Identification, Embodiment, Proteus effect.}
\begin{teaserfigure}
  \includegraphics[width=\textwidth]{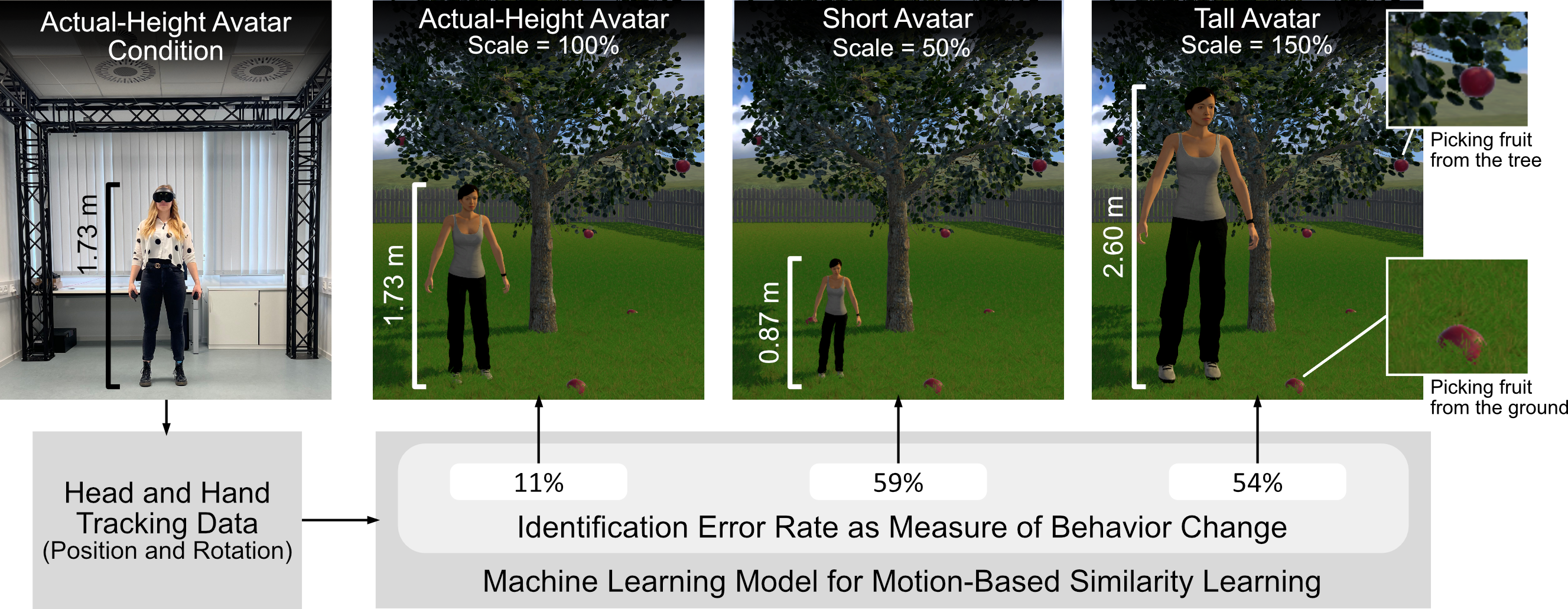}
  \caption{This figure shows the error rate of our motion-based similarity learning model when using motion data from a user of the condition with an avatar scaled to their actual height and measuring their behavior change in comparison to data from their i) actual-height avatar condition, ii) small avatar condition, or iii) large avatar condition. The model achieves low error rate when the appearance of the query data matches the reference data and high error rate when they do not match.}
  \label{fig:teaser}
\end{teaserfigure}


\maketitle
\pagestyle{plain} 
\section{Introduction}

Embodied interaction is a central characteristic of prominent applications of Virtual Reality (VR), eg, in social VR \cite{Freeman2020MyBodyMy}, therapy \cite{bartl2022avatar, doellinger2019vitras,turbyne2021review}, or exercise \cite{born2019embodiment}. 
Virtual embodiment is defined as the sense of having and controlling a virtual body in a virtual environment \cite{kilteni2012sense}.
An important topic in recent research is whether and how the digital representation of a user, the avatar, affects the user experience. 
For example, personalized photorealistic avatars have certain advantages in terms of user experience over other representations that do not match the appearance of the user \cite{fiedler2023embodiment,salagean2023meetingvirtualtwin,waltemate2018impact}. 

Interestingly, embodying an avatar that does not match the user's real appearance can induce changes in their behavior. The Proteus effect denotes a change of behavior caused by the avatar's appearance to conform to the perceived stereotype of the avatar \cite{yee2007proteus}. 
For example, users who embody taller avatars in a virtual negotiation task exhibited more confident behaviors than those with shorter avatars \cite{yee2007proteus}.
Recent work also discusses such behavior changes, suggesting that they might also be caused by what is called bodily affordances~\cite{oberdorfer2024proteus}. 
For example, a taller avatar allows for different interaction possibilities than a smaller one since the reach is higher, and users could interact differently, not based on appearance but on what they actually can do. 
Notably, both effects might act simultaneously. 
While changing the body height of an avatar changes the bodily affordances, it could additionally induce behavior change caused by the Proteus, as previous work showed \cite{yee2007proteus}.

However, independent of the concrete causality of a given behaviour change, the fact that a change occurs is interesting from various points of view, including objective assessment of intervention effectiveness, real-time feedback and adaptation, or just data collection for basic research of said effects and many more. 
So far, measuring behavior changes can be quite troublesome, and previous work primarily uses measures after the exposure like subjective questionnaires \cite{roth2020veq,peck:2021a}, association tests \cite{banakou2016racialbias}, or explicit behavioral measurements \cite{yee2009proteus}. 
Behavior changes typically lead users to modify their posture, gestures, and overall motion patterns. 
Still, objectively measuring behavior change in an implicit manner is underexplored.
There are attempts to measure behavior change caused by the avatar's appearance with objective quantitative methods \cite{kilteni2013drumming, oberdorfer2024proteus}.
However, they required complex additional hardware for motion capture and extensive analysis to find differences in movement specific to their use case, and they do operate after the exposure and finalized data collection. 

This paper proposes to utilize recent advancements in data-driven approaches to motion biometrics to detect behavior changes in-situ during exposure. 
We investigate whether and how a prominent machine-learning method initially developed for user identification can be used to continuously monitor and analyze motion data of head and hand to distinguish the user's motion and, subsequently \cite{miller2020behaviour}, a potential behavior change. 
Therefore, we investigate the research question:

\textbf{RQ:} Can we use generalizable data-driven analyses from off-the-shelf device motion data to measure behavior change when altering avatar appearance?


To address this question, we developed a data-driven motion analysis approach using a similarity learning model to measure behavior change.
To evaluate our method, we designed a study in which users embody in counterbalanced order: i) An avatar scaled to their actual body height, ii) a short avatar, or iii) a tall avatar in a virtual environment.  
During each condition, the motion of the head-mounted display (HMD) and controllers was tracked. 
After each VR exposure, we asked participants to estimate their avatar's body height and measured their sense of embodiment. 
We observed that participants' behavior change could be quantified, and their behavior varied between the different height avatars. 
This result highlights that behavior changes caused by the avatar's appearance are quantifiable with motion-based similarity learning models. 
We compared our approach to subjective questionnaires and task-specific motion-based analysis to validate it.
While they show similar results, our approach holds distinct advantages: 1) In-situ measurement without additional user input, 2) generalizable and scalable motion analysis for various use cases, 3) user-specific analysis on the individual level, 4) with a trained model, users can be added and evaluated in real time to study how avatar changes affect behavior.

\section{Related Work}

Altering the avatar's appearance can lead to behavioral changes in the user ~\cite{latoschik2017effect}.
The Proteus effect is a highly researched phenomenon in VR~\cite{banakou2013proteus,coesel2025hidden,li2014wiimyselfsize,lin2021exercisingyoungavatar,lin2021exercisingsixpack,peck2013puttingyourself,yee2007proteus}, describing how changes in the visual representation of one's avatar can alter the user's behavior to conform to the avatar's appearance stereotypes ~\cite{yee2007proteus}.
Previous work has shown that a higher sense of embodiment, and especially a higher perceived virtual body ownership, can lead to a stronger Proteus effect ~\cite{beaudoin2020impact,kilteni2013drumming}.

\subsection{Proteus Effect and Bodily Affordances}
There are numerous examples of the Proteus effect demonstrating the impact across various contexts, such as influencing negotiation strategies \cite{yee2007proteus}, interpersonal attitudes \cite{peck2013puttingyourself}, and physical behaviors like during drumming \cite{kilteni2013drumming}, walking \cite{reinhard2020acting}, or workouts \cite{li2014wiimyselfsize,lin2021exercisingyoungavatar,lin2021exercisingsixpack}.
Users adapt their behavior to conform to the stereotype they associate with the avatar based on its appearance.
Questionnaires \cite{peck:2021a, roth2020veq}, association tests \cite{banakou2016racialbias}, or explicit behavioral measurements like choosing an exercise weight \cite{mal2023proteus} are used to measure the Proteus effect. 
Previous work also tried to objectively measure the behavior change caused by the Proteus effect \cite{kilteni2013drumming}.
Kilteni et al. \cite{kilteni2013drumming} considered the positional data of the whole upper body with a motion capture system and showed that the behavior changed significantly when altering the avatar's appearance. 
For their analysis, they applied a principal components analysis (PCA).
In addition, they examined the frequency of hand movements by calculating the ratio of detected movement peaks between the baseline and altered avatar appearance conditions.

Bodily affordances of one's avatar can also significantly influence user behavior \cite{oberdorfer2024proteus}. 
Oberdörfer et al. \cite{oberdorfer2024proteus} defined bodily affordances as "the cause for a behavioral change in accordance with the expected constraints of the avatar". 
In their work, they showed that users with different embodiments have different body poses. 
For this, they compared the positional and rotational data with motion-captured data, showing significantly different body movements between the avatar appearances they used.

Therefore, a behavioral change caused by an altered avatar appearance, whether it is the Proteus effect or bodily affordances, can be quantified using extensive analysis of the motion pattern of the users when their motion is captured \cite{kilteni2013drumming, oberdorfer2024proteus}.
Nevertheless, previous work specifically uses complex motion capture and analysis of task-specific movement to measure the behavior change caused by the avatar's appearance.
The motion capture system required additional hardware, and users had to wear specific body-tracking suits, making this measurement rather intrusive.
Hence, there is a research gap in measuring behavior change caused by the altered avatar appearance that is generalizable, scalable, non-intrusive, and not constricted or proven to work only in a specific domain.

\subsection{Motion-Based User Identification in VR}
Recent work has demonstrated that machine-learning approaches can analyze motion patterns in such a detailed way that users can be differentiated and identified solely on their motion data of head and hands in various applications \cite{miller2020behaviour, nair2023identification, rack2023alyx, rack2024motion}. 

Rogers et al. \cite{rogersApproachUserIdentification2015} were the first to apply motion data for user identification, successfully distinguishing between 20 individuals. 
Miller et al. \cite{miller2020behaviour} expanded this concept, demonstrating scalability by accurately identifying 511 users. 
Rack et al. \cite{rack2023alyx} further advanced this domain by utilizing deep learning techniques on a dataset comprising 71 users playing Half-Life Alyx. 
Nair et al. \cite{nair2023identification} showed remarkable scalability, identifying up to 50,000 users playing Beat Saber with 94.33\% accuracy using only 100 seconds of motion data. 
However, these studies primarily used classification methods, which are limited to recognizing only individuals seen during training, as they directly predict a known class label corresponding to a previously observed user.  

In contrast, feature-distance or similarity learning methods have the advantage of being able to identify users not included in the training data.  
These methods return an embedding—a high-di\-men\-sional vector that abstractly represents a user’s motion patterns.
Previous research has demonstrated the feasibility of identifying individuals based on such embeddings.
Li et al. \cite{liWhoseMoveIt2016} demonstrated that feature-distance methods could verify individuals with an accuracy of 95.57\% based on head motion while listening to music. 
Miller et al. \cite{millerUsingSiameseNeural2021} showed the potential of similarity learning by identifying 41 individuals performing ball-rolling tasks across different VR systems. 
Rack et al. \cite{rackVersatileUserIdentification2024} used a dataset with 71 users playing Half-Life Alyx to confirm that similarity learning methods can be used to identify users. 
Rack et al. \cite{rack2024motion} also showed that similarity learning yields comparable results to the feature distance approach for uniform ball-throwing motion sequences, but it significantly surpasses these models in more intricate tasks such as handwritten in-air signatures.

Similarity learning has proven effective in reliably identifying users and distinguishing between user behavior based on motion data.  
We aim to leverage this property to detect behavioral changes in users that are provoked by different avatar appearances, indicated by a degradation in identification performance.

\section{Method}
\label{sec:method}
Current research often employs questionnaires \cite{roth2020veq}, behavioural tasks after exposure \cite{yee2009proteus}, and non-learned motion analysis \cite{kilteni2013drumming, oberdorfer2024proteus} to detect behavioral changes.
Questionnaires and behavioral tasks, however, have the limitation that they can only be analyzed retrospectively, making it difficult to determine when and how a behavioral change occurred, and they do not allow for user-specific analysis.
Moreover, they lack the granularity to specify the nature of behavioral differences.
However, non-learned motion analysis requires a thorough understanding of the study context and careful consideration of which motion metrics might differ between experimental conditions.
Often, several metrics must be tested before identifying one that reliably indicates a behavioral change.
While this allows for more precise insights into user motion or behavior, it also demands a deeper analysis of the sub-movements under investigation and requires the tracking of those.

We address these challenge by proposing a novel methodology that allows for user-specific behavioral change analysis without requiring detailed task-specific knowledge or additional tracking solutions.
This methodology is intended to support real-time detection of behavioral changes based on general motion characteristics of off-the-shelf devices, thereby enabling timely interventions during the study.
To realize this, we built on a system from the machine learning domain that identifies users based on the motion data from HMDs and controllers.
Our core idea is to investigate whether identification accuracy declines when a user's behavior changes.
We have summarised in \autoref{tab:behavior_comparison} which measures offer support for which features.

We introduce our novel approach for non-intrusive, objective, in-situ measurement of behavioral change.
We describe its design, as well as the preprocessing and training procedure for the similarity learning model.
We utilize this model to detect behavioral changes by analyzing declines in identification accuracy.
Specifically, we compute an identification error rate defined as:

\begin{equation}\label{eq:ier}
\textit{identification error rate} = 1 - \textit{identification accuracy}.
\end{equation}

This reveals how much the user's motion differs.
For example, an identification error rate of 0\% indicates that the user's motion is identical to the original, showing no difference between the ground truth and the new motion.

\begin{table}[htb]
\centering
\small
\begin{tabular}{l|l|l|l}
  \multicolumn{1}{l}{} 
    & \multicolumn{1}{l}{\rotcell{\textbf{Questionnaires}}} 
    & \multicolumn{1}{l}{\rotcell{\textbf{Non-learned motion analysis}}} 
    & \multicolumn{1}{l}{\rotcell{\textbf{ML-Based Identification Error}}} \\
  \hline
  \textbf{Retrospective evaluation of BC}        & Yes & Yes & Yes \\
  \textbf{In-situ detection of BC}               & No  & Yes & Yes \\
  \textbf{User-specific effect assessment of BC} & No  & No  & Yes \\
  \textbf{Omit analysis of BC on measured factors} & No & No  & Yes \\
  \textbf{Real-time BC detection}                & No  & No  & Yes \\
  \hline
\end{tabular}
\caption{Comparison between Questionnaires, Non-learned motion analysis, and ML-based identification error rate in the context of measuring behavior change (BC).}
\label{tab:behavior_comparison}
\end{table}

\subsection{Similarity Learning Method}
\label{sec:user_identification}
Our approach is a similarity learning model based on Deep Metric Learning, which maps the input data onto an embedding space, in this case, motion sequences.
In this space, embeddings with the same labels have minimal distances, while embeddings with different labels have larger distances.
Embeddings are multidimensional vectors, and the distances between these vectors represent relationships between motion sequences. 
Our model is trained to compute embeddings such that motion sequences from the same user result in embeddings with small distances. 
In contrast, motion sequences from different users produce embeddings with larger distances.
The primary advantage of this approach is that the trained model generalizes to new users who are not present during training or validation. 
Rather than directly learning individual user identities, the model learns the differences between motion patterns and encodes these differences in the embedding space.

For the evaluation, we split the data into two subsets: a reference set and a query set.
The reference data serves as the baseline and consists of motion sequences with known user labels used to compute reference embeddings. 
The query data contains motion sequences derived from new sequences we intend to investigate, and their embeddings are computed as query embeddings.
Each query embedding is compared to the reference embeddings to evaluate user identification performance. 
The user is identified as the one associated with the reference embedding with the smallest distance to the query embedding. 

To investigate how much a user's movement differs between the reference and query embeddings, we select multiple random query embeddings and check whether the correct user is identified with the reference embeddings.
From this, we can calculate the identification accuracy, which allows us to calculate the identification error rate as specified by eq.~\eqref{eq:ier}.


\subsection{Preprocessing of the Data}
\label{sec:preprocessing}
We preprocessed the raw data for training and evaluation by removing irrelevant information (i.e., noise), a step that was shown to improve the training process, following methods developed by Rack et al.\cite{rackComparisonDataEncodings2022}.
As part of this preprocessing, we resample the motion data at different frame rates to identify the optimal frame rate for the model during the hyperparameter search. 
The data is then transformed into body-relative. 
This makes each position and rotation of the frames relative to the HMD local coordinate System. 
This removes the HMD position because it is always the origin (0,0,0). 
After that, we transformed it into body-relative-velocity (BRV) encoding to removing irrelevant factors such as the user's position or orientation within the scene. 
This transformation prevents overfitting and ensures that the model focuses on relevant behavior. 
After these preprocessing steps, the input sequence is represented by 18 features per frame: ($rot-x, rot-y, rot-z, rot-w$) for the HMD and ($pos-x, pos-y, pos-z, rot-x, rot-y, rot-z, rot-w$) for each controller (left and right).

\begin{figure*}[]
  \centering  \includegraphics[width=\linewidth]{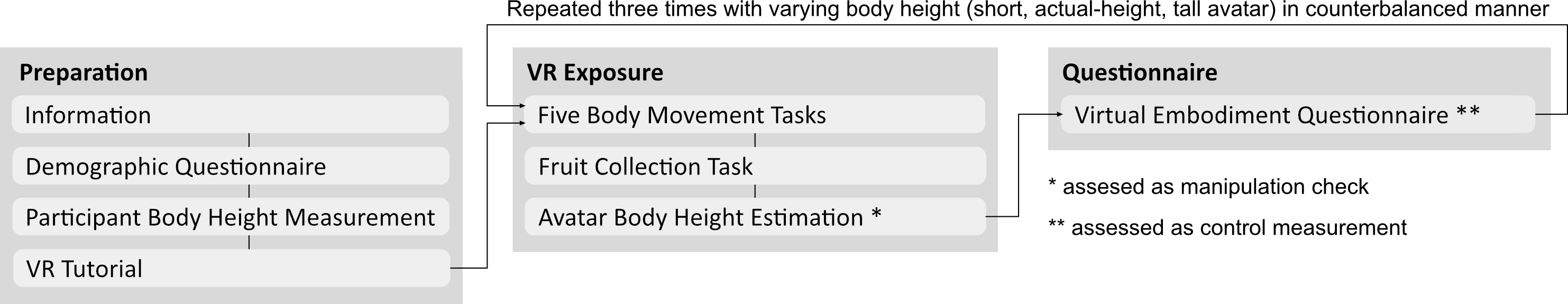}
  \caption{Outline of the user study procedure.}
  \label{fig:procedure}
\end{figure*}

\subsection{Architecture and Training of the Model}
\label{sec:model}
We use a transformer-based architecture for the similarity-learning model.
This design builds on prior work \cite{rack2024motion, liebersUnderstandingUserIdentification2021, nair2023identification}, which has demonstrated that deep learning approaches are highly effective for user identification.
Initially, we use the preprocessed data and feed it into the model as motion sequences. 
A motion sequence is a series of, for example, 600 consecutive frames, each consisting of 18 features. 
The first layer in our architecture is a Dropout layer, which randomly removes individual frames during training. 
This improves the model's robustness to variations in the input data. 
Next, a GRU layer processes the sequence data with tunable hyperparameters such as the number of layers or hidden size. 
The output is then passed to a TransformerEncoder~\cite{vaswaniAttentionAllYou}, which includes further hyperparameters such as the number of layers, attention heads, or the feedforward dimension. 
Finally, the data passes through two fully connected hidden layers that generate the final output: the embedding. 

 The implementation is done in Python using the PyTorch Lightning\footnote{https://lightning.ai} and PyTorch Metric Learning\footnote{https://kevinmusgrave.github.io/pytorch-metric-learning/} libraries, details can be found in the published code and dataset. Our architecture includes a large number of hyperparameters, which were optimized using "Sweeps" function of the Weights and Biases service\footnote{https://wandb.ai}.
After hyperparameter tuning, during which we trained multiple models with different hyperparameter combinations, we selected the best-performing model based on training metrics. 
This selection was guided by performance on the validation dataset, using R-Precision and Precision@1 as evaluation metrics.
These metrics capture different aspects of performance. 
They are computed by selecting random query embeddings from the validation dataset and comparing them to the reference embeddings. 
R-Precision measures the proportion of reference embeddings from the same user that are located in the immediate neighborhood of the query embedding. 
Precision@1 indicates how often the closest reference embedding to a given query embedding belongs to the same user.
The selected model was then used for evaluation.
In this evaluation phase, motion sequences from unseen users are passed through the model to compute embeddings directly via a forward pass.

\section{Evaluation} 
\label{sec:evaluation}

To evaluate our method, we used a $1\times3$ within-subject design with a counterbalanced order, manipulating the scale of the virtual environment to induce the illusion of an altered avatar's body height.
Participants embodied avatars of three different body heights: (1) a small avatar, scaled to 50\% of the participant's body height; (2) an actual-height avatar, which had the participant's body height; a tall avatar, with 150\% of the participant's body height. 
We use different measurement approaches, including questionnaires, task-spersific motion analyses, and our machine-learning approach explained in \autoref{sec:method}.

\subsection{Participants}
22 participants (7 male), aged 19 to 63 years ($M = 26.91$, $SD = 12.27$) took part in the study. 
Seven were undergraduates earning course credits, and 15 received monetary compensation. 
All participants fulfilled the criteria: (1) regular or corrected vision and hearing, (2) a minimum of ten years of local language proficiency, and (3) no color blindness.

\subsection{System Description}

\label{sec:system-description}
The VR system was developed using Unity 2021.3.32f1~LTS and integrated hardware via the Oculus XR plugin (version~3.3.0). 
It ran on a Windows 10 workstation with a Meta Quest Pro head-mounted display (HMD).
For body tracking, we used Captury's markerless system\footnote{https://captury.com/resources/} with eight FLIR Blackfly S BFS-PGE-16S2C RGB cameras mounted on the ceiling, capturing body motion at \unithertz{70}. 
The cameras were connected to an Ubuntu workstation running Captury Live version~261b, and body motion was streamed to the VR system. 
We retargeted the body pose to the avatar and merged it with the head tracking of the VR system using Unity's avatar animation system and a custom retargeting script.

We used the avatars from the Rocketbox Library~\cite{gonzalesfranco2020rocketboxavatars}. The avatars' appearances were gender-matched and dressed in casual sportswear to align with the study's context, as participants were intended to relax in their orchards. Initially, we planned to manipulate the avatar's body height. However, scaling the avatar's body height caused significant inaccuracies in mapping participants' motion to the avatar. Instead, we scaled the virtual environment to create the illusion of an altered avatar's body height. For the condition where the avatar appeared taller, the environment was reduced by 50\%, and for the condition where the avatar appeared shorter, it was enlarged by 50\%.  
The virtual environment resembled an orchard with a garden house, fruit trees, scattered fruit, and a fence defining the walkable area (see \autoref{fig:teaser}). Instructions were given through the HMD speakers and were displayed on a virtual board. 

\subsection{Fruit Collection Task}

Participants were tasked with collecting fruit. 
They had to pick up five fruits from trees and five fruits from the ground and store them in a virtual backpack. 
A display showed the number of collected fruits. 
Participants could walk around and use the "Grab-the-Air" locomotion technique for further navigation \cite{mapes1995grabmove}. 
This method was chosen because the virtual orchard was larger than the physical lab space, making natural walking infeasible. 
Fruit was scattered on the ground and distributed at different heights on the trees so that each participant could complete the task in each condition and collect enough fruit. 
For each avatar's body height condition, the task of picking the fruit from the trees was feasible. 
The participants pressed the trigger button on a controller to grab the fruit. 
A backpack function was implemented to count collected fruit and make it disappear when the participant placed it behind their shoulder.

\subsection{Procedure}

\label{sec:study-procedure}
Our study followed a standardized procedure, detailed in \autoref{fig:procedure}, with an average duration of \unitminute{45} and each exposure lasting about \unitminute{7}. 
First, the participants' body height was measured, and they completed a demographic questionnaire. 
Before the first VR exposure, they received instructions on using the HMD and controllers. 
During the whole VR exposure, participants received instructions via pre-recorded voice commands and written on a virtual board.
At the start of each VR exposure, participants entered a dark virtual preparation environment for eye tests, avatar calibration (duration: \unitsecond{30}), and a brief tutorial on interaction and navigation within the virtual environment. 
Next, the HMD display was blacked out, a pre-programmed test sequence was launched, and the participants entered the orchard environment, where their generic, gender-matched virtual avatar became visible. 
In each VR exposure, participants completed five body movement tasks \cite{waltemate2018impact}, each lasting for \unitsecond{20} before starting the main task: the fruit collection task. 
A UI showed the number of collected fruit. 
To collect the fruit, the participants had to stretch or bend down, depending on the height of the fruit and their avatar's body height, and then they placed the fruit in their virtual backpack behind their shoulders. 
After completing the task, the participants were asked to estimate the avatar's body height. 
Following each VR exposure, the sense of embodiment was assessed. 
This procedure was repeated three times, varying the avatar's body height, before compensation was provided.

\subsection{Measurements}
This section presents the three main measurement approaches of our study: a machine learning method to detect behavioral changes, a task-specific analysis of motion data, and questionnaires for subjective measurement and control variables.

\subsubsection{Machine Learning Method}

\label{sec:machine-learning-method}

As described in \autoref{sec:method}, we use this machine learning approach to calculate the identification error rate, which indicates behavioral changes.
We utilized all motion data from each condition in which users embodied their respective avatars at one of the three heights. 
As detailed in chapter \autoref{sec:study-procedure}, this encompasses the interval from entering the orchard environment until the avatar's height estimation is complete.
Out of the 22 recorded users, we excluded one due to corrupted movement data. The remaining 21 users were preprocessed as explained in \autoref{sec:preprocessing} and subsequently divided into three subsets: nine users for training, five for validation, and seven for evaluation.
During training, the model was provided with data from all conditions and optimized solely to accurately identify users.
We trained over 500 models with varying hyperparameters. 
Throughout the hyperparameter search, the model demonstrating the highest precision@1 in the validation dataset was selected for evaluation.
The hyperparameters, their search space, and configurations used in our final network are documented in the published code (during the review process, all resources can be found in the Supplementary Material). 

To evaluate the selected model, we computed the identification error rate across different conditions. 
We evaluated the model with various query and reference embeddings as described in Section \ref{sec:user_identification}.
In our dataset, reference embeddings were computed using motion sequences from a specific study condition. 
Query embeddings were generated using motion sequences from either the same or different conditions.
The motion sequences each have a duration of 40 seconds, represented by 600 frames.
Through the resampling process, our data was 15 frames per second (FPS).

\subsubsection{Task Specific Motion Data Evaluation}
\label{sec:analysis-movement-data}
We analyzed the motion data to determine whether differences between the three conditions could already be identified without applying machine-learning techniques. 
For this, we used the same raw motion data as in our machine-learning approach. 
We aimed to investigate whether users showed different motions when embodying different avatars in our different conditions. 
To achieve this, several analyses, listed below, were conducted. 
User 3 was excluded from these calculations due to a calibration error that led to a wrong logging of the motion data in one of the conditions. 
However, this user was still included in the analysis and in the machine-learning model, as the embodiment worked as intended, and the pre-processing steps mitigated the impact of this error.
Since our implementation is based on Unity, we adopt its coordinate system, where the forward direction corresponds to the $z$-axis, the right direction to the $x$-axis, and the up direction to the $y$-axis.

\paragraph{Hand over Head}
\label{sec:hand-over-head}
The analysis involved examining how frequently at least one hand was positioned above the head. 
For this analysis, we did not use the head height indicated by the HMD. 
Instead, we utilized the actual head height data collected from demographic measurements prior to the study. 
The reason we do not rely on the height displayed by the HMD is that it does not exactly represent the user's actual head position.
For the calculation, all participants were scaled to a uniform height of \unitcmeter{170} to standardize our data. 
The coordinate system was then converted into body-relative coordinates, which were calculated as follows: The user coordinates were rotated around the Y-axis to ensure that all participants were oriented forward, and the X- and Z-axes were centered relative to the HMD position. 
For the analysis itself, the number of frames in which at least one hand passed the standardized height threshold of \unitcmeter{170} was calculated.

\paragraph{Looking Up or Down}
\label{sec:looking-up-down}
For this analysis, we looked at the degree to which users looked up and down in the respective conditions.
Therefore, we converted the participant data into body-relative coordinates, as described in \autoref{sec:hand-over-head}. 
Then, we calculated the angle around the X-axis using the head rotation data to identify whether participants looked up or down. 
In this way, we calculated the mean and standard deviation of the head rotation angle per user for each condition.

\subsubsection{Questionnaires}
Participants completed questionnaires using LimeSurvey~4 and verbally answered in-experience questions. 
We ensured questionnaire accuracy by using validated translations or conducting back-and-forth translations for questions in the local language. 
To measure the sense of embodiment, we used the Virtual Embodiment Questionnaire (VEQ)~\cite{roth2020veq} with its scales of virtual body ownership, agency, and change on a 7-point Likert scale (1--7). 
While virtual body ownership and agency measure the sense of embodiment as defined by Kilteni et al. \cite{kilteni2012sense}, change measures a perceived change in the own body scheme through embodying the avatar, a potential instrument to measure the Proteus effect \cite{roth2020veq}. 
To measure the perceived body height of the avatar, participants were asked at the end of each VR exposure \enquote{How do you estimate the body height of your virtual body?}.

\section{Results}
\label{sec:results}
In this section, we present the outcomes of our experimental analyses, structured into assessments of subjective experiences of the user study, task-specific motion data, and performance evaluation of our similarity learning model. 
We used R version 2022.07.2\footnote{https://www.R-project.org/} for statistical analysis.

\subsection{Performance Evaluation of the Model}
\begin{figure}[htb]
 \centering
\includegraphics[width=0.8\linewidth]{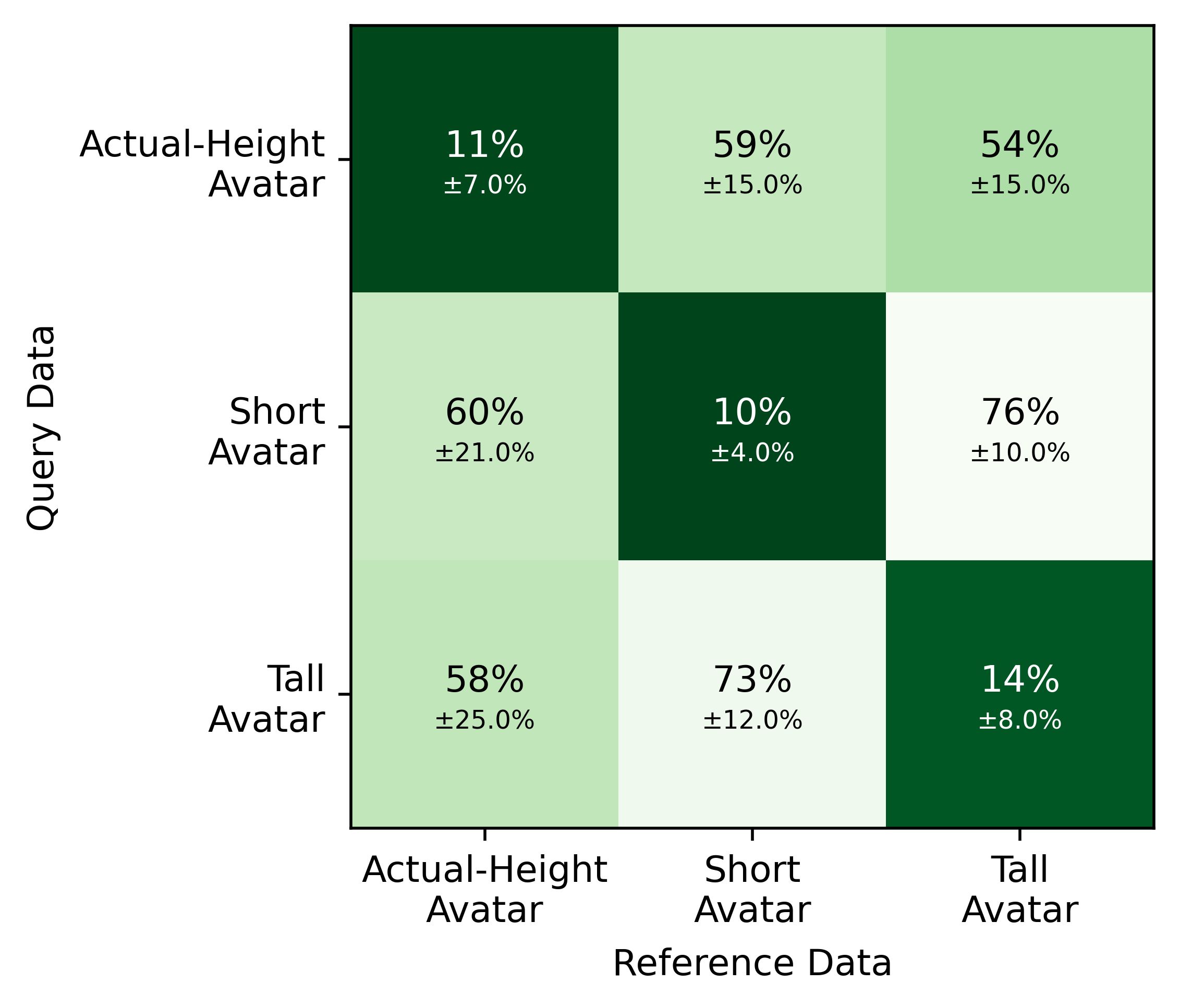}
  \caption{This figure shows the identification error rate in the different conditions, indicating with a higher deviation, a higher measured behavior change. The query data used is shown in the row, and the reference data is in the respective column.}
  \label{fig:matrix}
\end{figure}

In the following, we present the results of our method for determining whether user behavior changes when embodying different avatars in the three conditions.
We utilized a similarity learning model for this task as described in \autoref{sec:machine-learning-method}.
The results are summarized in \autoref{fig:matrix}.
When the reference and query data originate from the same condition, we calculate the error rate from one condition within the same condition so the avatar's body height remains constant. 
This is represented by the diagonal in \autoref{fig:matrix}, where error rate ranges from \unitpercent{10} to \unitpercent{14}, with a maximum significance of \unitpercent{8}.
On the other hand, if the reference and query data are from different conditions, we calculate the identification error rate from one condition to another.
Then users will have a one- or two-level difference in body height. 
One- or two-level difference in body height refers to the difference in scale, i.e., the short to actual-height avatar has a one-level difference, and the short to-tall avatar has a two-level difference.
In this case, the identification error rate increases significantly to a range of \unitpercent{54} and \unitpercent{76}  (i.e., the non-diagonal entries in \autoref{fig:matrix}). 

We validated these results on the test dataset using bootstrapping to assess consistency.
The mean of bootstrapped accuracies showed a maximum difference of \unitpercent{2} to the original values, with a standard deviation of \unitpercent{3.2}.
We also evaluated the model’s robustness using three different random seeds. 
Across these seeds, the accuracy varied by at most \unitpercent{38}.

\subsubsection{Contrast Test between Accuracies}
\label{sec:contrast-test-between-accuracies}
We used weighted Wilcoxon contrast tests to determine whether significant differences exist between the accuracy values of different groups.
For this purpose, we categorized the identification error rates into three distinct groups based on the degree of similarity between the query and reference data with respect to the avatar's body height.
The first group includes all identification error rates in which the query and reference data originated from the same condition, meaning the avatar's body height was identical.
The second group includes all accuracies involving a one-level body height difference between the avatars, such as comparisons between short and actual-height or actual-height and tall.
In these cases, either the query or the reference sample comes from the actual-height avatar condition, while the other comes from either the short or tall avatar condition.
The third group encompasses identification error rates in which the avatar's body height differed by two levels. For instance, the query and reference data originate from the short and tall avatar conditions, respectively.
Since each group had different numbers of accuracies, we used weighted tests when computing the contrasts.
The first contrast, between Group 1 and Group 2, showed a significant difference ($z(6)=2.37, p=.0180, r = 0.89$), resulting in higher accuracy for Group 2 (M=57.8\%, SD=18.41) than Group 1 (M=11.8\%, SD=6.55).
The second contrast, between Group 1 and Group 3, showed the same values ($z(6)=2.37, p=.018, r = 0.89$), resulting in higher accuracy for Group 3 (M=74.6, SD=10.98) than Group 1 (M=11.8\%, SD=6.55).
The third contrast, between Group 2 and Group 3, showed a significant difference ($z(6)=2.03, p=.0425, r = 0.77$), resulting in higher accuracy for Group 3 (M=74.6, SD=10.98) than Group 2 (M=57.8\%, SD=18.41).

\subsection{Task Specific Motion Data Evaluation}
As described in \autoref{sec:analysis-movement-data}, we analyzed the motion data to determine whether there were differences between the three conditions.

\begin{figure*}[htb]
  \centering
\includegraphics[width=0.9\linewidth]{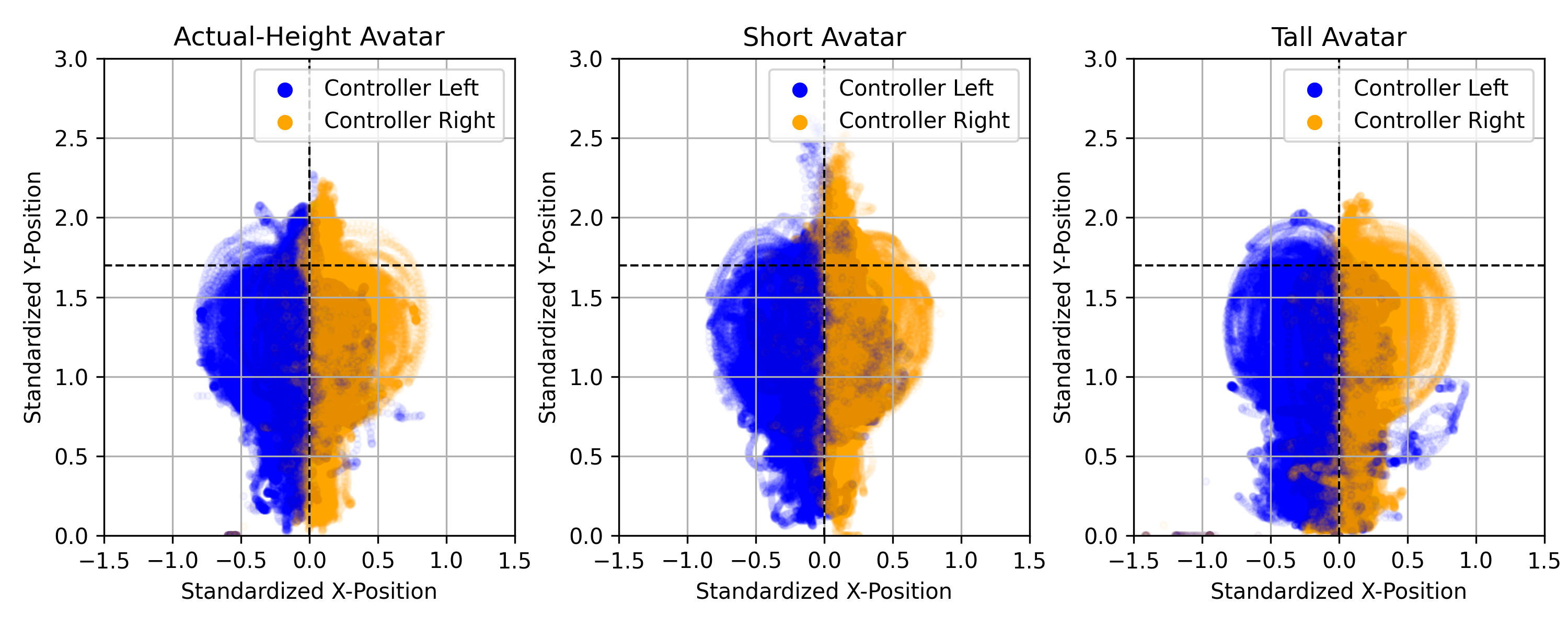}
  \caption{The left figure shows identification error rates across conditions; larger deviations indicate greater behavioral changes. 
The three figures on the right show aggregated controller positions across all participants for the three conditions. 
All avatars were rescaled to a standard height of 170 cm (dashed horizontal line); the vertical dashed line at \(x = 0\) marks the body midline.
}
  \label{fig:test_grafik_in_one_line}
\end{figure*}

\subsubsection{Hand over Head}
As described in \autoref{sec:hand-over-head}, we analyzed how often users raised their hands above their heads.
The hand positions of all users under each condition are shown in \autoref{fig:test_grafik_in_one_line}.
For the actual-height avatar, at least one hand was above the head in \unitpercent{2.98} of the recorded positions.
While embodying the short avatar, this occurred in \unitpercent{4.11} of the positions, and for the tall avatar in \unitpercent{1.50}, calculated across all users.
We calculated a $1\times3$ non-parametric Friedman test due to violated normal distribution to test whether the number of frames in which one hand was above the head differed significantly between conditions and found a significant main effect ($\chi^2$(2)$ = 21.9, p < .001, W = 0.548$). 
Post-hoc pairwise Wilcoxon tests with Bonferroni correction showed significantly higher values for the actual-height avatar compared to the tall avatar ($W (20,20) = 208, p < .001, r = 0.856$) and for the short avatar compared to the tall avatar ($W (20,20) = 206, p < .001, r = 0.839$).

\subsubsection{Head Orientation}
As described in \autoref{sec:looking-up-down}, we analyzed the extent to which users looked up or down in VR across the different avatar height conditions.
For the actual-height avatar, participants looked downward with a mean of \unitdegree{14.95} (SD = \unitdegree{7.12}).
While embodying the small avatar, the mean downward gaze was \unitdegree{9.48} (SD = \unitdegree{6.81}).
For the tall avatar, the mean downward gaze increased to \unitdegree{21.11} (SD = \unitdegree{9.09}).
We conducted a $1\times3$ ANOVA with Greenhouse-Geisser sphericity correction to test whether the gaze direction differed significantly between the conditions, which revealed a main effect for the condition ($F(1.25,23.72) = 0.624$, $p < .001, \eta_{p}^{2} = 0.592$).
Post-hoc pairwise T-tests with Bonferroni correction showed that the amount of downward gaze was significantly higher for tall compared to average-height avatars ($t(19) = 4.23$, $p = .001, d = 0.945$) and compared to small avatars ($t(19) = 5.64$, $p < .001, d = 0.1.26$). Furthermore, the amount of downward gaze was significantly higher for actual-height avatars compared to short avatars ($t(19) = -5.47$, $p < .001, d = -1.22$)

\subsection{Subjective Experiences of the User Study}
\label{sec:results-user-study}
All descriptive data are in \autoref{tab:results} and plotted in \autoref{fig:test_grafik_in_one_line}. 
We calculated $1\times3$ non-parametric Friedman tests due to violated normal distribution.

As expected, there were only tendencies for a main effect between the conditions for virtual body ownership ($\chi^2$(2)$ = 5.83, p = .054, W = 0.132$), no main effect for agency ($\chi^2$(2)$ = 1.61, p = .448, W = 0.037$), but a main effect for change ($\chi^2$(2)$ = 10.21, p = .006, W = 0.232$). 
Post-hoc pairwise Wilcoxon tests with Bonferroni correction showed significantly lower values for the actual-height avatar compared to the tall avatar ($W (22) = 30.50, p = .03, r = 0.549$) and for the actual-height avatar compared to the short avatar ($W (22) = 141.00, p = .007, r = 0.659$). 

As a manipulation check, we compared the relative deviation in estimated avatar body height across conditions ($\Delta =$ EstimatedAvatarBodyHeight $-$ ParticipantBodyHeight), where negative values indicate underestimation and positive values indicate overestimation. 
We found a main effect ($\chi^2$(2)$ = 36.10, p < .001$). 
Post-hoc tests revealed significant differences between all groups: the tall avatar was estimated larger than the actual-height avatar ($W (22) = 189, p < .001, r = 0.803$), and the short avatar ($W (22) = 231, p < .001, r = 0.853$); the actual-height avatar was estimated taller than the short avatar ($W (22) = 2, p < .001, r = 0.794$). 

\begin{table*}[]
  \centering
  \label{tab:results}
\begin{tabular}{lcccl}
\toprule
                                   & Actual-Height Avatar ($M$ ($SD$)) & Short Avatar ($M$ ($SD$)) & Tall Avatar ($M$ ($SD$)) & Test statistics                  \\ \midrule
VEQ Body Ownership         & $4.13$ ($1.77$)                   & $4.10$ ($1.61$)           & $3.70$ ($1.45$)          & $\chi^2$(2)$ = 5.83, p = .054$  \\
VEQ Agency                         & $5.47$ ($1.03$)                   & $5.50$ ($1.18$)           & $5.42$ ($0.85$)          & $\chi^2$(2)$ = 1.61, p = .448$  \\
\textbf{VEQ Change}                & $3.56$ ($1.46$)                   & $4.25$ ($1.49$)           & $4.69$ ($2.01$)          & $\chi^2$(2)$ = 10.21, p = .006$ \\
\textbf{$\Delta$ Body Height (cm)} & $-6.45$ ($11.15$)                 & $-40.05$ ($33.33$)        & $54.50$ ($47.73$)        & $\chi^2$(2)$ = 36.10, p < .001$   \\ 
  \bottomrule
\end{tabular}
  \caption{The table presents the descriptive statistics for each experimental condition, including the Virtual Embodiment Questionnaire (VEQ) \cite{roth2020veq} and relative deviation $\Delta$ in body height estimation as described in \autoref{sec:results-user-study}. Variables with a significant main effect are highlighted in bold.}
\end{table*}

\section{Discussion}

We introduced a novel approach to measure behavior change in-situ during VR exposure using a generalized motion analysis based on a similarity learning model. 
This method enables unobtrusive in-situ data-driven behavioral assessment without additional post-experiment input, offering a valuable alternative to subjective self-report measures and advantages of other motion analysis approaches.
To validate the method, we designed a user study in which participants appeared to embody an avatar based on the participant's body height in three conditions to induce behavior change caused by the avatar appearance: i) actual-height avatar, ii) short avatar, or iii) tall avatar, in a counterbalanced order.
Our manipulation check, which assessed participants' estimates of the avatar's body height, confirmed the effectiveness of our manipulation since participants rated the shorter or taller avatar as significantly shorter or taller than the actual-height avatar.
Notably, the avatar height manipulation did not significantly affect participants’ sense of \textit{Virtual Body Ownership} or \textit{Agency}, suggesting that the sense of embodiment \cite{kilteni2012sense} was maintained across all conditions. This ensures that observed behavioral changes can be attributed to the manipulation.

\subsection{Implications for Measuring Behavior Changes}

For our method, we used a similarity learning model. 
The model was only trained to identify users, not the different conditions.
Our results show that the identification error rate is low when the query and reference data match the same condition, so when the avatar's body height is identical.  
However, the identification error rate increased significantly when comparing motion data across different avatar body height conditions.  
This reveals an interesting pattern.  
When we examine the identification error rate for comparisons involving a one-level body height difference—such as between short and actual-height or actual-height and tall avatars—the identification error rate was higher.  
Additionally, the identification error rate increased even more when the body height difference was two levels (e.g., comparing short and tall avatars). 
Therefore, the identification error rate increases gradually and significantly with increasing avatar-height differences.  
Additionally, our statistical results show that the identification error rate increased for each user, showing that the different avatar appearances resulted in behavior change for every user. 
Therefore, we can affirmatively answer our research question:
Generalizable data-driven analyses from off-the-shelf device motion data can measure behavior change when altering avatar appearance.

To validate these findings, we compare our method first to the results of the task-specific motion analysis and second to the questionnaires:
Since participants had to reach with their hands to the trees, we calculated the number of frames in which the participants put their hands above their heads and the head rotation, that is, how often the participants' head orientation is up or down.
Those who looked down the most were participants during the tall condition, which is only logical since most of the virtual environment was beneath them. 
While head rotation metrics revealed differences between all conditions, hand movement differences were only observed between two comparisons. 
These metrics allow for measuring the behavior change as it has been shown similarly by previous work \cite{kilteni2013drumming, oberdorfer2024proteus}.
However, our approach of a motion-based similarity learning model holds a distinct advantage as it does not require predefined movement features and instead captures generalized behavioral shifts. 
It further requires less data since we can use motion data provided by off-the-shelf devices, and we do not need additional motion capture systems.
This makes it more robust to task variability and allows it to identify subtle, distributed changes in behavior.
Consequently, the model offers the advantage of improved user-specific analyses, making it easier to identify individual behavioral changes. 
In contrast, task-specific motion analysis is less robust, resulting in greater variability between users, which complicates conducting user-specific analyses.

Second, the \textit{Change} scale of the VEQ showed that participants experienced a significantly greater change in their body schema when embodying the short or tall avatar than when embodying the actual-height avatar. 
Compared to the questionnaire, our model operates unobtrusively and in situ, requiring no additional participant effort, recall, or adaptation of the procedure, as there is no step for answering a questionnaire. 
While the VEQ’s Change scale identified body schema shifts between the actual-height condition and the short and tall conditions, it did not distinguish all three. 
In contrast, our method captured behavior differences more comprehensively between all the conditions.
Additionally, as the model is user-specific, it enables behavior change detection at the individual level, allowing us to identify whether the behavior change was apparent for every user and not only for some. 
Allowing for a more in-depth analysis of whether the avatar's appearance induces behavior change.
In summary, our deep metric similarity learning approach has the following advantages:
\begin{enumerate}
\item[1:] 
In-situ measurement without additional user input. It captures behavior change directly from motion data without interrupting or altering the user's experience. 
Our approach utilizes motion data already captured by standard XR systems. 
We leverage the tracking of head and hand tracking from off-the-shelf devices without additional hardware or sensor integration. 
This allows seamless integration into immersive environments and makes the analysis process more accessible and less intrusive.

\item[2:] 
The similarity learning model is not bound to specific tasks or predefined movement features. It instead learns generalizable representations of motion patterns, making it adaptable to a wide range of use cases beyond the presented study. Prior work \cite{nair2023identification} has demonstrated that such models can scale to datasets involving tens of thousands of users. 

\item[3:] 
The similarity learning approach operates on a user-specific level. It can detect behavior changes within individual users, allowing for personalized insights and an adaptive system. 

\item[4:]
As soon as we have a small data foundation from users and a trained similarity learning model, we can add users without retraining the model, even during exposure in real-time.
Using the reference data of a user's avatar appearance, we can investigate whether an altered avatar appearance causes a change in behavior during runtime.
\end{enumerate}

\subsection{Limitations and Future Work}

Our study uses a small and homogeneous sample for motion-based similarity learning.
Additionally, participants had short exposure time in VR, compared to prior studies that typically rely on larger datasets \cite{ miller2020behaviour, nair2023identification, rack2023alyx}.
This limitation poses challenges for both training and evaluating the model. 
A larger and more diverse dataset, incorporating more users and extended VR interaction time per user, would likely yield even more generalizable and meaningful results. 
We have to verify and validate our approach with a larger sample size and alternative avatar appearance changes, like manipulating the age \cite{banakou2013proteus}, the skin texture of the avatar \cite{banakou2016racialbias}, the clothes \cite{mal2023proteus}, or the sex \cite{slater2010first}.
Since such visual modifications can systematically influence user motion, as shown in our results, they have implications not only for behavior analysis but also for motion-based user authentication and identification.  
In social VR, where users frequently adapt their avatars to express identity, self-perception, or situational context \cite{Freeman2021BodyAvatarMe, Freeman2020MyBodyMy, bimberg2024environment, Maloney2020AnonymityvsFamiliarity}, even subtle appearance changes may affect motion consistency.  
Future work on motion-based authentication and identification should account for these avatar-induced behavioral variations when designing and training their models.  
Further, future work can gain in-depth insights when our approach is carried out over the entire course of the study, as it can be applied even during XR exposure.
This could provide a more accurate measurement of behavioral changes over time.
Additionally, an intriguing direction for future work would be to explore whether the direct distance between query and reference embeddings—analogous to its application in motion-password authentication \cite{rack2024motion}—could serve as a more granular quantitative indicator of behavioral change.

\section{Conclusion}

This work uses a motion-based similarity learning model to present a novel, non-intrusive method for in-vitro measurement of behavior change in XR. 
We demonstrate that avatar appearance (avatar height) significantly alters user motion, and our model reliably captures these changes across varying avatar height conditions. 
Our approach enables user-specific detection of behavioral shifts induced by altered avatar appearance without relying on subjective input or task-specific features. 
By bridging the gap between data-driven motion analysis and embodiment research, this method offers a scalable and generalizable tool for measuring behavior change caused by bodily affordances and the Proteus effect.
We contribute a behavioral metric for off-the-shelf XR systems and future user studies on avatar appearance and behavior change.
In future work, we will apply our measure to other avatar appearance manipulations to validate our method.

\bibliographystyle{ACM-Reference-Format}
\bibliography{references}






\end{document}